\documentclass[12pt,preprint]{aastex}

\shorttitle{The White Dwarf Luminosity Function}
\shortauthors{Harris et al.}

\begin{document}

\title{The White Dwarf Luminosity Function from SDSS Imaging Data}

\author{Hugh C. Harris\altaffilmark{1,11},
Jeffrey A. Munn\altaffilmark{1},
Mukremin Kilic\altaffilmark{2},
James Liebert\altaffilmark{3},
Kurtis A. Williams\altaffilmark{3},
Ted von Hippel\altaffilmark{2},
Stephen E. Levine\altaffilmark{1},
David G. Monet\altaffilmark{1},
Daniel J. Eisenstein\altaffilmark{3},
S.J. Kleinman\altaffilmark{4},
T.S. Metcalfe\altaffilmark{5},
Atsuko Nitta\altaffilmark{4}, 
D.E. Winget\altaffilmark{2},
J. Brinkmann\altaffilmark{4},
Masataka Fukugita\altaffilmark{6},
G.R. Knapp\altaffilmark{7},
Robert H. Lupton\altaffilmark{7},
J. Allyn Smith\altaffilmark{8,9},
Donald P. Schneider\altaffilmark{10} }

\altaffiltext{1}{U.S. Naval Observatory, 10391 W. Naval Observatory Rd.,
 Flagstaff, AZ 86001-8521}
\altaffiltext{2}{Department of Astronomy, Univ. of Texas, Austin, TX 78712}
\altaffiltext{3}{Steward Observatory, Univ. of Arizona, 933 N. Cherry Ave.,
Tucson, AZ 85721}
\altaffiltext{4}{Apache Point Observatory, PO Box 59, Sunspot, NM 88349-0059}
\altaffiltext{5}{High Altitude Observatory, NCAR, PO Box 3000,
Boulder, CO 80307}
\altaffiltext{6}{Institute for Cosmic Ray Research, University of Tokyo,
 5-1-5 Kashiwa, Kashiwa City, Chiba 277-8582, Japan}
\altaffiltext{7}{Princeton Univ. Observatory, Peyton Hall, Princeton, NJ 08544}
\altaffiltext{8}{Los Alamos National Laboratory, P.O. Box 1663, Los Alamos,
 NM 87545}
\altaffiltext{9}{Department of Physics and Astronomy, Univ. of Wyoming,
 P.O. Box 3905, Laramie, WY 82071}
\altaffiltext{10}{Department of Astronomy and Astrophysics,
 The Pennsylvania State Univ., 525 Davey Laboratory, University Park,
 PA 16802}
\altaffiltext{11}{hch@nofs.navy.mil}

\begin{abstract}
A sample of white dwarfs is selected from SDSS DR3 imaging data
using their reduced proper motions, based on improved
proper motions from SDSS plus USNO-B combined data.
Numerous SDSS and followup spectra (Kilic et al. 2005)
are used to quantify completeness and contamination
of the sample;  kinematic models are used to understand
and correct for velocity-dependent selection biases.
A luminosity function is constructed
covering the range $7 < M_{\rm bol} < 16$, and its
sensitivity to various assumptions and selection limits is
discussed.  The white dwarf luminosity function based on 6000
stars is remarkably smooth, and rises nearly monotonically to
$M_{\rm bol} = 15.3$.  It then drops abruptly, although
the small number of low-luminosity stars in the sample
and their unknown atmospheric composition prevent quantitative
conclusions about this decline.  Stars are identified that may
have high tangential velocities, and a preliminary luminosity
function is constructed for them.
\end{abstract}


\section{Introduction}

The white dwarf luminosity function (WDLF) is important in
providing a record of the star formation history and the
age of each of the components of the Galaxy in the solar
neighborhood.  Historically the hot end of the WDLF has been
determined by selection of blue stars (e.g. 
Fleming, Liebert, \& Green 1986;  Vennes et al. 2002;
Liebert, Bergeron, \& Holberg 2005 [LBH]),
while the WDLF at medium and cool WD temperatures often
has been determined by selection that includes proper motion
(e.g. Liebert, Dahn, \& Monet 1988 [LDM];  Boyle 1989;  Evans 1992;
Oswalt et al. 1995;  Knox, Hawkins, \& Hambly 1999).
The faintness and rarity of WDs requires that large areas of the
sky be covered in order to acquire a large WD sample and to determine
the WDLF accurately.  Photographic survey data generally
provide poor-quality photometry that hampers WD selection.
The desirable goals of high-accuracy photometry and astrometry
over wide areas of the sky motivate us to use digital surveys
such as the Sloan Digital Sky Survey (SDSS).

The SDSS (York et al. 2000; Stoughton et al. 2002; Gunn et al. 1998)
is obtaining images of the sky in a survey area around the
North Galactic Cap and in other smaller areas.  From these images,
calibrated photometry in the five $ugriz$ filters (Fukugita et al.
1996; Lupton et al. 2001; Hogg et al. 2001;  Smith et al. 2002;
Ivez\'ic et al. 2004) and star-galaxy image classification are
obtained for all detected objects.  SDSS then obtains spectra
of many objects, primarily galaxies and QSOs and a small fraction
of stars.  For purposes of studying WDs, spectra are
obtained of essentially all hot WDs with $T > 22000$~K,
and of many medium-temperature WDs with $8000 < T < 22000$~K
(Harris et al. 2003; Kleinman et al. 2004), but of very few
cool WDs with $4000 < T < 8000$~K, until they reach the rare
ultracool temperatures $T < 4000$~K (Gates et al. 2004).
This incomplete sampling of the spectra of WDs with cool and
intermediate temperatures, just in the temperature range where
the luminosity function is near its maximum, is a result of their
colors overlapping the far more numerous F, G, and K main-sequence
stars.  An attempt to distinguish cool WDs from field stars by
combining SDSS photometry with another narrow-band magnitude
was only partially successful (Kilic et al. 2004).
Therefore, neither spectroscopic nor photometric selection
of WDs in SDSS are useful approaches for studying the WDLF
at intermediate and cool temperatures.

Instead, identification of WDs with cool and intermediate
temperatures in SDSS imaging data is feasible by including
proper motions.  This paper uses proper motions derived from
combined SDSS and USNO-B astrometry, along with SDSS photometry,
to define a sample of WDs selected using reduced proper motions.
The WDLF based on this sample is presented, and the important
factors and limitations of the data are discussed.
The selection procedure is described in Sec. 2.
Factors that should be accounted for in constructing the
WDLF are discussed in Sec. 3, including estimates of the
completeness and contamination of the WD selection.
In Sec. 4, initial results are presented for the WDLF
based on SDSS Data Release 3 (Abazajian et al. 2005).

\section{Selection of White Dwarfs by Reduced Proper Motion}

Identification of WDs in SDSS imaging data is
done here using the reduced proper motion (RPM) diagram.
The RPM, in g magnitude, for example, is defined as
\begin{displaymath}
H_g = g + 5 {\rm log} \mu + 5 = M_g + 5 {\rm log} V_{\rm tan} - 3.379
\end{displaymath}
where $\mu$ is the proper motion in arcsec yr$^{-1}$
and $V_{\rm tan}$ is the tangential velocity in km~s$^{-1}$.
The RPM provides a distance-independent tool for separating stars
with very different absolute magnitudes.  White dwarfs are
typically 5--7 mag less luminous than subdwarfs of the same color,
so they have RPM values typically 3--4 mag fainter than subdwarfs
after the approximately three times greater velocity dispersion
of subdwarfs is accounted for.  The RPM diagram has been used
in previous studies to identify WDs (Evans 1992; Knox et al. 1999;
Oppenheimer et al. 2001; Nelson et al. 2002)
and subdwarfs (e.g. Digby et al. 2003).
The accurate SDSS colors help in the separation of different
stellar populations.

Calculating values of RPM requires measuring proper motions,
and SDSS is, in general, only a single-epoch survey.
However, matching SDSS objects with a catalog made from
sky survey plates immediately makes it possible to measure
the proper motions for most objects bright enough to be
detected and included in the plate-based catalog.
This paper uses the improved proper motion catalog
given by Munn et al. (2004), extended to include the area of
5282 deg$^2$ of sky in the SDSS Third Data Release
(Abazajian et al. 2005).
It is based on matches to the all-sky USNO-B Catalog
(Monet et al. 2003), but it includes proper motions that are
improved over those in USNO-B both by including the SDSS positions
(Pier et al. 2003) and by correcting the
sky-survey positions to an inertial system using galaxies --
see Munn et al. (2004) for details.  The minimum proper motion
of 20 mas~yr$^{-1}$ used by Munn et al. is also used here,
eliminating nearly all QSOs and many nondegenerate stars.
The resulting catalog has errors in proper motion of 3-4
mas~yr$^{-1}$ and is nearly 90\% complete at $g = 19.5$;  we
discuss the effects of these errors and incompleteness in Sec. 3.4.

In a companion paper, Kilic et al. (2005) present the RPM diagram
(their Figures 1 and 6) based on the Munn et al. (2004) catalog.
Kilic et al.\ use SDSS spectroscopy, as well as follow-up spectroscopy
of regions of the RPM diagram expected to be occupied by cool WDs
but not observed spectroscopically by SDSS; they confirm that WDs occupy
a locus in the RPM diagram, cleanly separated from most subdwarfs,
and that samples of WDs can be defined using the RPM diagram
with contamination by subdwarfs and QSOs of only a few percent.
We use that result to create statistically complete catalogs of WDs.
A region of the RPM diagram is defined by a curve showing the
cooling track of WDs with pure hydrogen atmospheres, using the models
of Bergeron et al. (1995), for some assumed tangential velocity.
A sample of WDs is defined by taking all stars below and blueward of
one of the model curves (for example, the 
V$_{\rm tan}$ = 30 km~s$^{-1}$ curve for pure-H atmosphere WDs).
Using V$_{\rm tan}$ cutoffs of 20, 30, and 40 km s$^{-1}$ yields
samples of 7116, 6000, and 4501 stars, respectively.  The extent of
contamination and incompleteness, the procedure used to correct for
interstellar reddening, and other issues relevant to the
luminosity function calculation, are discussed in Section 3.
The list of stars selected with V$_{\rm tan} > 20$ km~s$^{-1}$
is given in a file available by anonymous ftp from
ftp.nofs.navy.mil/pub/outgoing/hch/tabdata.dat.

\section{Construction of a Luminosity Function}

The luminosity function is calculated based on distances
to each star derived from photometry (see Sec. 3.1 and 3.2).
The high quality SDSS photometry greatly aids this procedure.
The effects of the disk scale height and completeness
are discussed in Sec. 3.3 and 3.4.
In Sec. 3.5, we examine and apply corrections for contamination.
The problem of accounting for the different properties of hydrogen
and helium atmospheres of cool WDs is discussed in Sec. 3.6.

\subsection{WD+M Binaries}

Some WD candidates have unresolved dM companions
(Raymond et al. 2003;  Kleinman et al. 2004;  Smolci\'c et al. 2004).
They are identified in this paper in the following way.
When all five SDSS magnitudes are fit to WD model colors, stars
that have a (poor) best fit with $\chi^2 > 20$ and that have
significant excess residual $u$ and $z$ flux from the best
fit model are designated as having a significant cool companion.
(The dM companion contributes to the red filters, but the initial
fit tries to match the composite flux distribution, so the
compromise WD model is too cool and residual excess flux is
observed in both $u$ and $z$ filters.)
These stars are shown as filled circles in Figure 1.
About 2.5\% of the sample selected by proper motion in this paper
are found this way to be composite.
For these stars, the remaining analysis is carried out using
only the $ugr$ magnitudes to fit for the temperature,
luminosity, distance, etc.  Omitting the $i$ and $z$ magnitudes
minimizes the effect of the cool companions, but does not
entirely eliminate it.  In practice, those WDs with fainter
cool companions are analyzed approximately correctly with this
procedure, a few WDs with companions of similar brightness at $i$
have their luminosities and distances underestimated,
and WDs with bright companions are lost entirely from our sample.

\subsection{Photometric Distances}

All five SDSS magnitudes (three for binaries in the previous section)
for each star are fit to WD models with hydrogen atmospheres to
determine the temperature, absolute magnitude, and distance of each
star.  The models of Bergeron et al. (1995) are used, with SDSS colors
kindly calculated by Bergeron (2001, private communication),
assuming log $g = 8$, and accounting for
reddening\footnote{The SDSS database gives the total interstellar
absorption and reddening along the line of sight for each star,
determined from the reddening maps of Schlegel et al. (1998).
Because most of the WDs in this paper have distances within a
few hundred pc, they are affected by only a fraction of the total
absorption and reddening.  This fraction is determined as part
of the fitting for the distance of each star in an iterative
procedure.  We make the assumption that the absorption is zero
for stars with distances $<100$~pc, that it is the full absorption
for stars with distances from the Galactic plane $|z| > 250$~pc,
and that the absorption
varies linearly along the line of sight between these distances.
Because all stars in this paper are at relatively high Galactic
latitudes, the total (maximum) absorption and reddening is small:
for WDs in our sample, the median total absorption $A_g$ is 0.10,
the median total reddening $E(g-i)$ is 0.04.  Therefore, the errors
in the reddening using this procedure are tiny and have little
effect on the derived distances.
}. 
Approximate corrections from the observed
SDSS $u$ and $z$ magnitudes onto the AB flux system of -0.04 and
+0.02, respectively (Abazajian et al. 2004), are also included here.
Colors of the Bergeron et al. (1995) models are plotted in Figure 2.
They show that for warmer WDs the H/He composition has little effect,
provided the gravity is normal.
Incorrect luminosities and distances will be derived for WDs
with unusually low mass and gravity, believed to be about 10\%
of all WDs (Bergeron et al. 1992; Liebert et al. 2005),
and those with unusually high mass and gravity,
believed to be about 15\% of all WDs, as shown by the
dotted curves displaced from the normal-gravity curves in Fig. 2.
The actual percentages of WDs with abnormal mass in our sample
with magnitude and RPM limits will be somewhat different;
the percentages are fairly small and the effects are in
opposite directions, and we ignore them here.

For warmer WDs with $T > 5500$~K ($g-i < 0.7$), the $M_{\rm bol}$
and $M_g$ magnitudes of the pure hydrogen and pure helium models
vs. $g-i$ in Fig. 2 are nearly coincident.  Therefore, the error
in using pure hydrogen models for the minority of WDs that actually
have helium atmospheres is expected to be small.  Fitting the colors
of a pure helium model with the grid of pure hydrogen
models gives an error in $M_{\rm bol}$ that varies between
+0.04 and -0.17 over the temperature range 5500--20000~K.
For example, a pure helium WD at 10000~K has $M_{\rm bol} = 11.85$,
but when fit with pure hydrogen models is assigned 10090~K and
$M_{\rm bol} = 11.78$.  Thus, some will go into a luminosity bin
higher than they should, and be assigned a distance (and $V_{max}$
value) too small.  The overall effect on the LF will be small.
Of course, the few WDs with unusual abundances (DQ and DZ spectral
types with strong absorption features) will have incorrect results,
but these stars are a small fraction of the total (Harris et al.
2003; Kleinman et al. 2004).

For cool WDs with $T < 5500$~K ($g-i > 0.7$),
the H/He composition does significantly affect the color-absolute
magnitude relations in Fig. 2 and the derived distance and luminosity.
Independent data will be needed to utilize the SDSS sample fully.
Deep infrared photometry might provide the needed H/He classification,
for example.  Lacking infrared data, we make a weighted H/He
assignment for each cool star based on our knowledge of the
fraction of each type from studies in the literature.  We discuss
the effects of this procedure in Sec. 3.6.

Finally, of the seven published ``ultracool'' WDs with SDSS data
(Gates et al. 2004; Harris et al. 1999; 2001),
stars that are sufficiently cool that collision-induced absorption
has a significant effect on their SDSS colors,
three have $g < 19.5$ and can enter the sample.
Photometric distance estimates cannot be relied upon, however,
because the WD models do not match the observed colors adequately
(Gates et al. Figure 1).  Even worse, the distance is known
only for LHS~3250, and that star is apparently undermassive
and overluminous compared to the models that assume normal mass.
Therefore, we adopt the same $M_g$ and $M_{\rm bol}$ values for
all three stars, based on the trigonometric parallax of LHS~3250.
(See Gates et al. for further discussion of this issue.)
With this assumption, SDSS J0947 has $V_{\rm tan} = 19$ km~s$^{-1}$,
and drops out of the sample.  LHS~3250 and SDSS J1403 are the only
known ``ultracool'' WDs remaining in the sample.  LHS~3250 has a
known luminosity ($M_{\rm bol} = 16.2$);  the assumption for the
remainder of this paper is that SDSS J1403 has the same luminosity,
although it is likely to be somewhat cooler and could have a lower
luminosity (Gates et al. 2004).  These two stars become the
lowest luminosity WDs in the sample.


\subsection{Galactic Disk Scale Height}

The Galactic disk scale height of the WD population must be known
before deriving the WDLF, because luminous WDs near the SDSS magnitude
limit are at distances of several hundred pc from the Galactic plane
where their space density is significantly reduced.
Normally the scale height is determined by comparing the
observed WD numbers vs. $|z|$ distance to models of exponential
disks with different assumed scale heights, and finding the
maximum likelihood fit.  Some results from the literature are
$245 \pm 25$~pc (Green 1980),
$\sim300$~pc (Ishida et al. 1982),
$\sim270$~pc (Downes 1986),
$275 \pm 50$~pc (Boyle 1989), and
$260 \pm 40$~pc (Vennes et al. 2002).
Because these studies used relatively small WD samples
(typically 100--200 stars), all stars were included in a single fit,
and the models necessarily included an input LF to account for
different numbers of stars at different luminosities.
A consequence of this procedure is that the derived scale height
and the derived LF are more coupled than is desirable.

SDSS provides a large enough sample to derive a
scale height for {\it each} luminosity bin separately.
Thus the models to which the data are compared are each
calculated for a single luminosity and require no assumptions
about the LF.  The models are constructed by taking the
actual sky coverage of the survey area (for this paper,
the imaging area covered by SDSS DR3) with its actual
distribution of Galactic latitudes, taking the survey
magnitude limits ($15.0 < g < 19.5$) and $V_{\rm tan}$
limit (30 km~s$^{-1}$), and, for each
luminosity bin, calculating the expected numbers of stars
at each $|z|$ distance for different assumed values of scale
height.  A factor is also included to account for the
exclusion of stars with small $V_{\rm tan}$ at large distances
whose proper motions fall below the lower proper motion limit
of 20 mas~yr$^{-1}$.

Plots of the data for WDs in eight different luminosity ranges
are shown in Figure 3, where the data are fit to models with
scale heights of (top to bottom) $\infty$, 500, 400, 300, and 200 pc.
Table 1 shows the scale height and its uncertainty derived
for each luminosity bin.  The last three columns are a scale
factor needed to scale the observed number of stars to the
best fit model, the reduced chi-square for the best fit model,
and the number of stars in the luminosity bin.
It is unclear whether the trend shown in Table 1 of increasing
scale height toward lower luminosity WDs is a result of some
systematic effect in the selection or analysis of the WD sample,
or is real.  A real increase in scale height is expected from the
older age of the coolest WDs (coupled with the known heating of the
Galactic disk with time), and possibly from an increasing fraction
of thick-disk stars at cooler temperatures and lower luminosities.
However, the large increase in Table 1 at $M_{\rm bol} \sim 12$
is probably not realistic (see Sec. 5).
Taking the first four bins ($9 < M_{\rm bol} < 11$), where stars
extend to large enough distances to be most sensitive to the scale
height, but where thick-disk stars should still be a small minority,
the weighted mean value of the scale height is
340${^{+100}_{-70}}$ pc.  This result is somewhat larger than
that of other studies in the literature, though consistent
within the errors.  More accurate results will come from
analysis of hot WDs in SDSS, either a purely photometry-selected
sample or a pure spectroscopic sample, where the distances are
greater and the requirement for proper motion selection can
be dropped.  Below we adopt 250 pc for the
scale height for better comparison with those other studies.
The effect on the WDLF of adopting different values is shown in
Sec. 4, and further discussion of the increasing scale height
at low luminosities is in Sec. 5.

\subsection{Corrections for Completeness}

White dwarfs with low tangential velocites can have the same
values of RPM as main-sequence subdwarfs with high tangential
velocities, so the two types of stars overlap to some extent
in RPM.  In order to minimize contamination of the WD sample,
a minimum tangential velocity $V_{\rm min}$ is chosen.
Candidate WDs with $V_{\rm tan} < V_{\rm min}$ are rejected
from the sample, and a correction factor $\chi$, referred
to as the discovery fraction by Bahcall \& Casertano (1986),
is included in the WDLF calculation for the rejected stars.
Small values of $V_{\rm min}$ introduce more contamination
by subdwarfs, whereas high values reduce the contamination
at the expense of reducing the WD sample and making the
correction factor $\chi$ more uncertain.
Values of $V_{\rm min}$ near 25--30 km~s$^{-1}$ are optimal
for selecting WDs from SDSS data, because contamination is
tolerable and $\chi$ is still well determined.

A Monte Carlo kinematic model of the disk WD population
is used here to calculate $\chi$ for each $V_{\rm min}$.
The model is described  briefly by Liebert et al. (1999)
and is similar to models constructed by others
(e.g. Wood \& Oswalt 1998;  Hansen \& Liebert 2003;
Garc\'ia-Berro et al. 2004).
The important input parameters are the values of the
velocity ellipsoid for disk WDs in the solar neighborhood.
Here we use the values of dM stars in the solar neighborhood
from Reid et al. (1995): 43, 31, 25, and --22 km~s$^{-1}$ for
$\sigma_{\rm U}$, $\sigma_{\rm V}$, $\sigma_{\rm W}$, and
the lagging rotation velocity relative to the local standard
of rest (often called the asymmetric drift), respectively.
The resulting values of $\chi$ are given in Table 2,
where values are given for the full sky and for the part of
the sky included in our SDSS sample.  The two values are
slightly different because different lines of sight see
different projections of the velocity ellipsoid.
These values for the velocity ellipsoid are larger than are
often cited for the thin disk because the WD population
(especially at the low luminosities included in this paper)
is older than most constituents of the thin disk, but they
are smaller than for the thick disk (e.g. Reid 2005).
Experiments with other plausible parameters for the velocity
ellipsoid indicate the uncertainty in $\chi$ is about 10\%
of $(1-\chi)$.  Therefore the uncertainty has little effect
on the derived WDLF, even for $V_{\rm min}$ = 40 km~s$^{-1}$.

A second correction is needed for completeness of the
USNO-B Catalog and its match with SDSS objects.  This correction
depends on magnitude, color, and proper motion.  Both Monet
et al. (2003) and Munn et al. (2004) provide relevant data,
but here we apply some restrictions to reduce contamination
(see next section), so new completeness values have been
calculated.  We have run all stars in DR3 through the same
USNO-B matching and selection used for the WD selection,
except for the proper motion and tangential velocity limits.
We assume that the SDSS catalog of stars is complete,
and find the completeness of the USNO-B matches.
The large numbers of stars provide accurate measures of
completeness vs. magnitude and color:  the completeness varies
between 86--95\% for $g<19.25$, 83--94\% for $g<19.5$,
and 78--93\% for $g<19.75$, depending on color.
The completeness is also a function of proper motion (Gould 2003;
Munn et al. 2004), being essentially complete for proper motions
less than 40 mas yr$^{-1}$, dropping to about 70\% complete for
proper motions of 1000 mas yr$^{-1}$, in addition to the above
magnitude-dependent factor.
The correction at small proper motion (less than 100 mas yr$^{-1}$)
was derived by comparison with SDSS data along the celestial equator
in the fall sky, which has been observed many times over a period
of five years, and for which SDSS-based proper motions are thus
available.  For larger proper motions, the correction has been derived
by comparison with the rNLTT (Salim \& Gould 2003), under the assumption
that the rNLTT is itself 90\% complete (L\'epine \& Shara 2005).
The combined overall completeness for WDs in this sample is 
near 70\%, and the uncertainty is estimated to be of order 10\%.
The correction (vs. magnitude, color, and proper motion)
has been applied.

\subsection{Corrections for Contamination}

Contamination of the WD sample occurs in two ways.  First,
nondegenerate stars sometimes have a spurious large measured
proper motion, and thus acquire a large RPM that can scatter
the star into the WD sample.  Because F, G, and K main-sequence
stars are so numerous in SDSS, even a very small fraction of
bad proper motion measures (usually caused by confusion in
matching the five plate detection lists in USNO-B) can cause
noticeable contamination.  To minimize this source of contamination,
we require stars with $g-i > 0.12$ (where the stellar main sequence
becomes heavily populated in SDSS) to be matched on {\it all five}
USNO-B plates, plus matching the SDSS detection within 1 arcsec
at the epoch of the SDSS scan, and to have no other SDSS source
with $g < 22$ within 7 arcsec (see Kilic et al. 2005).
For stars with $g-i < 0.12$
(for which potential contaminants are far less numerous),
we relax these restrictions and accept objects detected on as
few as three USNO-B plates plus SDSS.  (However, in Sec. 5 we
discuss WDs found to have high tangential velocities.  For those
rare objects, the relaxed selection allows some contamination,
so in Sec. 5 we revert back to the tighter restrictions for
high-velocity WDs.)  These restrictions were chosen after
inspecting many WD candidates (and all cool WD candidates)
on digital scans of the sky survey plates used for USNO-B to
verify their proper motions (available at
http://www.nofs.navy.mil/data/fchpix).

A second source of contamination is nondegenerate stars with
correctly measured proper motions and with very high tangential
velocities that give RPM values in the WD selection region.
These can be either hot sdO and sdB stars or cooler main-sequence
subdwarfs.  The choice of $V_{\rm min}$ has a strong effect;
$V_{\rm min} = 15$ km~s$^{-1}$ allows a significant number of real
contaminating stars, whereas $V_{\rm min} = 30$ km~s$^{-1}$ 
eliminates most of them.  Only spectra can identify these
contaminants -- Kilic et al. (2005) show that this contamination is
at most a few percent, and probably much less.
Using these restrictions, we find contamination of 1--2\% for
most of the WD sample, exemplified by four SDSS QSOs that have
entered our sample.
The luminosity functions below
are corrected by these small factors.

\subsection{Cool White Dwarfs with Helium Atmospheres}

Figure 2 shows that for WDs with $g-i > 0.7$, corresponding to
temperatures cooler than 5500~K and $M_{\rm bol} > 14.5$, 
the distance and luminosity of a star with a helium-dominated
atmosphere will be underestimated if the pure hydrogen curve
is assumed to be correct and used for the photometric distance
determinations described in Sec. 3.2.  Spectra generally do not
help because, with the exception of a small fraction of stars
with carbon or calcium features, most cool WDs with either
hydrogen or helium atmospheres have featureless DC spectra.
A combination of accurate optical and infrared photometry usually
can distinguish their type (Bergeron et al. 2001), but infrared
photometry is not presently available for most of our sample.
(Only about 5\% of the WDs in this sample are detected in the 2MASS
Point Source Catalog.  We are in the process of obtaining infrared
photometry for many of the cool WDs in the sample.)
Therefore, for cool WDs, we will calculate
the distance and luminosity of each star under both hydrogen and
helium assumptions, and assign a fraction (depending on the color
of the star) to each type based on our knowledge of the H/He ratios
of cool WDs from the literature.  A similar procedure was followed by
Knox et al. (1999).  Here, however, each star's assignment is neither
total hydrogen vs. helium, nor random;  instead, each star can
contribute to the WDLF with both types through a weight W.

The fraction of hydrogen and helium atmosphere types that we adopt
for cool WDs is based on the following facts:  
few helium-atmosphere WDs are known with temperatures in the range
5000--6000~K (Bergeron et al. 2001), and only in the cooler half of
this temperature range does the composition matter for our purposes;
the fraction of known hydrogen-atmosphere WDs is 40--50\%
in the temperature range 4000--5000~K (Bergeron et al. 2001);
for temperatures below 4000~K, the few known WDs probably have
helium-dominated (but not pure helium) atmospheres
(Bergeron et al. 2001; Bergeron \& Leggett 2002; Gates et al. 2004). 
Because cool WDs with helium atmospheres are more luminous at a
given color (by 0.7 mag at $g-i = 1.3$, for example),
they can be seen at greater distances (1.4 times greater,
for the same example) than those with hydrogen atmospheres.
Therefore, in a magnitude-limited sample like SDSS,
in the reddest color range we should expect to find about three times
more helium-atmosphere stars, if the two types have equal
space densities in that color range.

In order to reach results that are consistent with the above facts,
the criteria given in Table 3 are adopted.  These criteria include a
most-likely H/He mix, labeled ``Best Fractions" in Table 3,
and two alternative mixes that favor more hydrogen types
(Alternative A) and more helium types (Alternative B).
All three mixes are used below to show the sensitivity of our results
to the H/He assumptions.  Ultimately, a determination of each star's
atmospheric composition is needed to help reduce the present ambiguity.

\section{White Dwarf Luminosity Function}

The LF is derived from the list of identified WDs by summing
the inverse volume of space in which each star potentially
would have been included within the sample limits,
calculating each luminosity bin separately.
The derived WDLF is shown in Figure 4.
A scale height of 250 pc and a tangential velocity limit
of 30 km~s$^{-1}$ have been adopted.

The result at the bright end is in acceptable agreement with
results from the Palomar-Green (PG) survey.  The latest
analysis of DA stars only in the PG survey (Liebert et al. 2005)
is shown in Fig. 4, where their $M_V$ space densities have been
converted to $M_{\rm bol}$ space densities.  Accounting for
non-DA stars, as was done by Leggett et al. (1998), will raise
the densities from the PG survey by a small amount.
The PG survey gives densities in agreement with SDSS results
for $7 < M_{\rm bol} < 10$, but smaller than SDSS
by factors of 2--3 for $10 < M_{\rm bol} < 13$ or
$11.3 < M_{V} < 13.2$.  This is exactly the range where Liebert
et al. (2005) suspect the PG-survey results are incomplete.
Both curves are plotted here, assuming a scale height of 250 pc.

The result at the faint end is in excellent agreement with
that from the local sample selected from the LHS Survey
(Leggett et al. 1998).  It agrees with two other similar proper
motion selected studies, Boyle (1989) and Knox et al. (1999).
It does not agree with some other results that have suggested
the LHS sample is incomplete.  We return to this point below.
The drop in density at the faint end is not yet a secure result,
and we also discuss this important issue below.
The integral of the curve in Fig. 4 gives a space density
of WDs in the solar neighborhood of 0.0046$\pm$0.0005~pc$^{-3}$.
The assumptions made here about the scale height and the fraction
of He-atmosphere stars affect this result, but not drastically
(see next paragraphs).  The space density is somewhat higher
than found by Leggett et al. (1998), 0.0034~pc$^{-3}$,
but lower than found by Holberg et al. (2002), 0.0050~pc$^{-3}$.

The effect of varying the tangential velocity selection limit
$V_{\rm lim}$ is shown in Figure 5.  Ideally, the correction factor
$\chi$ would correctly account for stars missed below the limit,
and these curves would coincide.  The higher densities derived
using $V_{\rm lim} = 20$ km~s$^{-1}$ suggests that some contamination
is entering the sample more than has been accounted for.
Contamination is expected to be a factor for the faintest two bins,
where the number of WDs is dropping but the space density of
cool subdwarfs is continuing to rise, and they enter the WD sample
at bright RPM values when a small $V_{\rm lim}$ value is used.

The effect of using different values for the disk scale height
is shown in Figure 6.  This factor affects only the bright end
of the WDLF.  As discussed in Secs. 3.3 and 5, a value of 300 pc
or more is suggested from the SDSS imaging data in this paper.
A more accurate analysis should be possible using hotter WDs
in SDSS, either from photometric selection or from spectra,
because they reach distances well beyond one scale height.

The different possible mixtures of hydrogen-dominated and
helium-dominated atmospheres given in Table 3 result in the
different WDLF results shown in Figure 7.  As expected, adopting
a higher fraction of hydrogen-dominated stars (Alternative A)
results in more stars being assigned fainter luminosities and gives
a less pronounced drop and a more extended faint tail to the LF.
There are very few stars in the faintest luminosity bins, and we
currently know nothing about the H/He type of each individual star,
beyond the likely types discussed in Sec. 3.6.
Furthermore, lacking a parallax for each star, the possibility
of an abnormal mass adds to the uncertainty in any conclusions from
these data.  Particularly for the two ``ultracool'' stars that are
assigned $M_{\rm bol} = 16.2$ (Sec. 3.2), one has a known luminosity,
but the other does not and could well be lower luminosity.
Therefore, the present results must be considered preliminary.
We can conclude that the SDSS data indicate a rapid decline in
the WDLF at $M_{\rm bol} > 15.4$, a result that supports conclusions
from the LHS sample.  However, the shape of the drop, the exact
luminosity of the drop, and the extent of the faint tail are
currently better determined from the LHS sample than from SDSS data.

\section{Discussion}

The WDLF derived here and shown in Fig. 4 is remarkably smooth
and featureless.  The only noticeable feature in the range
$8 < M_{\rm bol} < 15$ is the small plateau near $M_{\rm bol} = 10.5$.
A plot with finer binning is shown in Figure 8.  There it is seen
that the feature is actually a flattening at $M_{\rm bol} = 10.0-10.5$
and a drop at $M_{\rm bol} = 10.7$, corresponding to a temperature
boundary of 13,200~K ($g-i = -0.45$) for DA stars.
The ZZ~Ceti instability strip is at slightly cooler temperatures,
also shown on the figure, so it is probably not related to this
feature.  There is no significant contamination of the sample
by nondegenerate stars or QSOs near this color.
The feature could be an artifact caused by incorrect models of
stars near this temperature, which could in turn cause incorrect
photometric distances to be assigned and/or cause incorrect values
of $M_{\rm bol}$ to be assigned.  However, the models are believed
to be quite accurate in this temperature range.  No feature is
predicted (Fontaine et al. 2001) from a pause in WD cooling and dimming.
If the feature is real, it could reflect a nonuniform rate of
production of WDs in the Galactic disk.  Noh \& Scalo (1990)
modeled the effect bursts of star formation have on the WDLF,
and suggested that a tentative bump seen in the WDLF (Liebert et al.
1988) might be caused by such a burst 0.3~Gyr ago.  The cooling time
for a normal-mass WD to reach $M_{\rm bol} \sim 10.5$ is 0.3~Gyr,
and the main-sequence lifetime of a likely progenitor is
$\sim$2.5~Gyr, suggesting that a drop in star formation about
3~Gyr ago (after a burst or a long-duration higher rate of
star formation) might be the cause of the plateau seen here in Fig. 8.
Further models like those of Noh \& Scalo would help quantify this
suggestion.

The cool end of the WDLF from Fig. 4 is shown with finer binning
in Figure 9.  (The sawtooth shape in Fig. 9 is apparently a
statistical fluke, as we have checked for errors introduced by
improper binning.)  The abrupt drop in the LF occurs at
$M_{\rm bol} = 15.40$, although the exact shape of the drop
and the exact luminosity at which it occurs both depend on the
unknown H/He type of WDs in this sample, as discussed above.
Also evident in this plot is a rise in the LF at $M_{\rm bol}
= 15.1$--15.2.  This rise appears to be marginally significant.
It is at a luminosity consistent with the predicted onset of
convective coupling between the convective hydrogen atmosphere and
the degenerate core (Fontaine et al. 2001) causing temporary additional
release of internal energy and delayed cooling.
Observationally, the rise in Fig. 9, like the following drop, is
sensitive to the assumptions made here about H/He types and about
the luminosity of the coolest stars.  Further study of these stars
should help to verify the reality and exact shape of this feature.

The analysis in this paper has assumed that the WD population
has a distribution away from the Galactic plane that can be
described with a single exponential scale height.
The WD population is actually more complicated because the
mean age of hot, luminous WDs is systematically younger than
that of cool, low-luminosity WDs.\footnote{
The WD population is further complicated because a one-component
gaussian velocity ellipsoid and a single scale height are not
realistic at {\it any} luminosity.  For example, both Hansen
\& Liebert (2003) and Reid (2005) suggest reference models for
thin disk WDs that have smaller values for the velocity dispersion
and scale height than are adopted here.  However, both papers
discuss the excess of high-velocity WDs that are observed.
Understanding their origin in the thin or the thick disk
(Bergeron et al. 2005; Reid 2005) will require much more work.
}
Stars in the Galactic disk have been perturbed over time, so old
populations have higher velocity dispersions and larger scale heights
(e.g. Wielen 1977; Dehnen \& Binney 1998).
How much will this disk heating affect the present analysis?
Over the range of WD luminosities included in this paper,
their properties change as follows:
M$_{\rm bol}$ changes from 8 to 13 to 15;  the WD mass drops
from 0.63 to 0.59 to 0.58 $M_{\sun}$, and the WD cooling age
increases from 0.04 to 1.2 to 6.2 Gyr (Bergeron et al. 1995);
the initial progenitor mass drops from 1.9 to 1.5 to 1.4 $M_{\sun}$
(Ferrario et al. 2005);
the main-sequence lifetime increases from 2 to 3 to 4 Gyr;
the total age increases from 2 to 4 to 10 Gyr;
the vertical velocity dispersion $\sigma_{\rm W}$ increases
from 12 to 18 to 27 km~s$^{-1}$ (Nordstr\"om et al. 2004);
the scale height increases from 250 to 350 to roughly 500~pc
(Mihalas \& Binney 1981).
At the low-luminosity end of the range discussed here,
the increased velocity dispersion causes greater completeness
of the sample (a higher value of $\chi$), and the increased
scale height also causes greater completeness by a tiny amount
shown in Fig. 6.  The two factors would act to reduce the derived
WDLF slightly were we to include them in the analysis.  The changing
scale height with WD luminosity has a bigger influence
on the total surface density of WDs in the Galactic disk,
however.  This factor will be important when interpreting the
WDLF to derive the star formation history of the disk.

A small fraction of WDs have large values of $H_g$
and, if they are normal WDs, high tangential velocities.
The analysis described above gives 80 stars with
$V_{\rm tan} > 160$ km~s$^{-1}$, about 1\% of the total
WD population found here.  A few are WDs, but with spurious
proper motions -- applying a more restrictive selection
(requiring detection on all five plates in USNO-B and excluding
stars with another star nearby for all stars, not just redder stars;
see Sec. 3.5) gives 32 stars with $V_{\rm tan} > 160$ km~s$^{-1}$
(0.6\% of the total).  Table 2 shows that a population with
disk kinematics is expected to have less than 0.1\% with
tangential velocities this high.\footnote{
Note that a $V_{\rm tan}$ limit at a velocity lower than
140 km~s$^{-1}$ is expected to include significant numbers
of thin-disk WDs, according to Table 2, based on the velocity
ellipsoid used in Sec. 3.4.  Contamination by the thick-disk
WDs is expected to be even greater with a low velocity limit
(Reid 2005).
}
The excess found here can come from halo stars, thick-disk stars
(e.g. Reid 2005) or thin-disk ``runaway'' stars with anomalously
high velocities (e.g. Bergeron et al. 2005).
It also can come from thin-disk stars that have high mass and
therefore erroneously high derived values of luminosity, distance,
and $V_{\rm tan}$;  LHS~4033, from the Oppenheimer
(2001) sample, is an example of such a high-mass WD (Dahn et al.
2004).  Further data about these high-velocity candidates come from
SDSS spectra of 34 stars:  11 are non-DA WDs, mostly DB or DQ,
so they are still candidates for high-velocity stars, but we
have little information about their masses and exact luminosities;
23 are DA WDs, and determination of their gravities by fitting
the hydrogen line profiles in their spectra with the procedure
in Kleinman et al. (2004) shows that 21 have normal values of
log~$g\sim8.0$ and only two have log~$g>8.5$ and are probably
high-mass stars.  These spectra indicate that only a few of the
80 high-velocity candidates are actually high-mass stars instead.

If we make the assumption that all the high-velocity candidates
are halo stars, we can construct a luminosity function.
To account for the halo stars with $V_{\rm tan}$ below the
adopted cutoff, $\chi$ corrections must be applied;
Table 4 gives these values, calculated in the same way as
described in Sec. 3.4, and based on the halo velocity ellipsoid
from Morrison et al. (1990).
Figure 10 shows the luminosity function using the restrictive
sample of 32 high-velocity WDs for three different
$V_{\rm tan}$ cutoffs.  Table 5 gives this high-velocity sample.
The fact that the LF derived using the
higher cutoffs does not drop much suggests that the sample is
not dominated by a rapidly dropping high-velocity tail of the
thick disk or the thin disk.  However, the sample can have
significant numbers of ``runaway'' WDs from the thin disk or
thick disk that have acquired a velocity kick somehow,
and therefore do not follow the velocity distributions assumed
in the models.  Therefore, the LF in Figure 10 should be considered
as an upper limit for the halo until the actual composition of
the sample is better understood.
The shape of the LF (rising toward lower luminosities)
and the integrated space density ($4\times10^{-5}$ pc$^{-3}$)
are both consistent with models of an old, single-burst
population (e.g. Hansen 2001);  they are probably consistent
with an interpretation of the Oppenheimer (2001) sample where
some allowance is made for contamination by disk stars
(e.g. Torres et al. 2002).
However, the halo/thick-disk/thin-disk composition of this sample
must be better understood before using it to derive a space density
for halo WDs.

\acknowledgments

This research has made use of the USNOFS Image and Catalogue Archive
operated by the United States Naval Observatory, Flagstaff Station
(http://www.nofs.navy.mil/data/fchpix).

Funding for the creation and distribution of the SDSS Archive has
been provided by the Alfred P. Sloan Foundation, the Participating
Institutions, the National Aeronautics and Space Administration,
the National Science Foundation, the U.S. Department of Energy, the
Japanese Monbukagakusho, and the Max Planck Society. The SDSS Web
site is http://www.sdss.org/.

The SDSS is managed by the Astrophysical Research Consortium (ARC)
for the Participating Institutions. The Participating Institutions
are The University of Chicago, Fermilab, the Institute for
Advanced Study, the Japan Participation Group, The Johns Hopkins
University, the Korean Scientist Group, Los Alamos National Laboratory,
the Max-Planck-Institute for Astronomy (MPIA),
the Max-Planck-Institute for Astrophysics (MPA),
New Mexico State University, University of Pittsburgh,
Princeton University, the United States Naval Observatory,
and the University of Washington.


\clearpage

\begin{deluxetable}{lrrrrrr}
\tablecolumns{7}
\tablewidth{0pt}
\tablecaption{Galactic Disk Scale Height}
\tablehead{
\colhead{$M_{\rm bol}$}{\quad\quad\quad\quad} &
\multicolumn{3}{c}{Scale Height (pc)} &
\colhead{SF} &
\colhead{$\chi_{\nu}^2$} &
\colhead{N} \\
\colhead{} &
\colhead{Value} &
\colhead{Min.} &
\colhead{Max.} &
\colhead{} &
\colhead{} &
\colhead{} }
\startdata
 $\phn$9.0--$\phn$9.5 \dotfill & 326.& 185.& 780.& 127.3&  1.48& 424 \\
 $\phn$9.5--10.0 \dotfill & 319.& 185.& 700.& 78.1&  1.98&  586 \\
 10.0--10.5 \dotfill & 355.& 195.& $\infty$&  64.2&   0.73&  613 \\
 10.5--11.0 \dotfill & 354.& 190.& $\infty$&  62.6&   2.54&  547 \\
 11.0--11.5 \dotfill & 463.& 240.& $\infty$&  47.7&   2.69&  627 \\
 11.5--12.0 \dotfill & 751.& 280.& $\infty$&  36.3&   1.74&  623 \\
 12.0--12.5 \dotfill & 894.& 330.& $\infty$&  26.2&   1.62&  520 \\
 12.5--13.0 \dotfill & 862.& 230.& $\infty$&  18.5&   1.58&  428 \\
\enddata
\end{deluxetable}

\clearpage

\begin{deluxetable}{ccc}
\tablecolumns{3}
\tablewidth{0pt}
\tablecaption{Completeness for Disk WDs with $V_{\rm tan} > V_{\rm min}$}
\tablehead{
\colhead{$V_{\rm min}$} &
\colhead{$\chi$} &
\colhead{$\chi$} \\
\colhead{(km s$^{-1}$)} &
\colhead{All Sky} &
\colhead{SDSS DR3} }
\startdata
$\phn\phn$0 & 1.000 & 1.000 \\
$\phn\phn$5 & 0.990 & 0.991 \\
$\phn$10 & 0.961 & 0.965 \\
$\phn$15 & 0.914 & 0.924 \\
$\phn$20 & 0.852 & 0.869 \\
$\phn$25 & 0.780 & 0.802 \\
$\phn$30 & 0.700 & 0.726 \\
$\phn$35 & 0.616 & 0.646 \\
$\phn$40 & 0.532 & 0.565 \\
$\phn$45 & 0.452 & 0.486 \\
$\phn$50 & 0.377 & 0.410 \\
$\phn$60 & 0.248 & 0.276 \\
$\phn$70 & 0.154 & 0.174 \\
$\phn$80 & 0.090 & 0.102 \\
$\phn$90 & 0.049 & 0.056 \\
100 & 0.026 & 0.029 \\
110 & 0.014 & 0.015 \\
120 & 0.007 & 0.007 \\
130 & 0.004 & 0.004 \\
140 & 0.002 & 0.002 \\
\enddata
\end{deluxetable}

\clearpage

\begin{deluxetable}{cccccccccc}
\tabletypesize{\footnotesize}
\tablecolumns{10}
\tablewidth{0pt}
\tablecaption{Adopted Fractions of H- and He-Atmosphere Stars}
\tablehead{
\multicolumn{2}{c}{Initial, If Hydrogen} &
\multicolumn{2}{c}{Revised, If Helium} &
\multicolumn{2}{r}{Best Fractions} &
\multicolumn{2}{r}{Alternative A} &
\multicolumn{2}{r}{Alternative B} \\
\colhead{$M_{\rm bol}$} &
\colhead{$T_{\rm eff}$} &
\colhead{$M_{\rm bol}$} &
\colhead{$T_{\rm eff}$} &
\colhead{H} &
\colhead{He} &
\colhead{H} &
\colhead{He} &
\colhead{H} &
\colhead{He} }
\startdata
$<$15.20 & $>$4650 & \dots & \dots &
   1.0 & 0.0 & 1.0 & 0.0 & 1.0 & 0.0 \\
15.20--15.50& 4330--4650& 14.86--15.01& 4850--5020&
   0.5 & 0.5 & 0.8 & 0.2 & 0.3 & 0.7 \\
15.50--16.00& 3850--4330& 15.01--15.19& 4650--4850&
   0.2 & 0.8 & 0.5 & 0.5 & 0.1 & 0.9 \\
$>$16.00 & $<$3850 & $>$15.19 & $<$4650 &
   0.0 & 1.0 & 0.2 & 0.8 & 0.0 & 1.0 \\
\enddata
\end{deluxetable}

\clearpage

\begin{deluxetable}{ccc}
\tablecolumns{3}
\tablewidth{0pt}
\tablecaption{Completeness for Halo WDs with $V_{\rm tan} > V_{\rm min}$}
\tablehead{
\colhead{$V_{\rm min}$} &
\colhead{$\chi$} &
\colhead{$\chi$} \\
\colhead{(km s$^{-1}$)} &
\colhead{All Sky} &
\colhead{SDSS DR3} }
\startdata
$\phn\phn$0 & 1.000 & 1.000 \\
100 & 0.854 & 0.893 \\
120 & 0.792 & 0.842 \\
140 & 0.724 & 0.782 \\
160 & 0.648 & 0.712 \\
180 & 0.568 & 0.633 \\
200 & 0.488 & 0.551 \\
220 & 0.409 & 0.465 \\
240 & 0.333 & 0.383 \\
260 & 0.265 & 0.307 \\
280 & 0.206 & 0.240 \\
300 & 0.155 & 0.183 \\
\enddata
\end{deluxetable}

\clearpage

\begin{deluxetable}{rrrrrrrrcll}
\tabletypesize{\scriptsize}
\tablecolumns{11}
\tablewidth{0pt}
\tablecaption{WDs With Possible High Tangential Velocities}
\tablehead{
\colhead{RA} &
\colhead{Dec} &
\colhead{$g$} &
\colhead{$g-i$} &
\colhead{$M_{\rm bol}$} &
\colhead{D} &
\colhead{$\mu$(RA)} &
\colhead{$\mu$(Dec)} &
\colhead{$V_{\rm tan}$} &
\colhead{Spectral} &
\colhead{Notes} \\
\multicolumn{2}{c}{(degrees)} &
\colhead{} &
\colhead{} &
\colhead{} &
\colhead{(pc)} &
\multicolumn{2}{c}{(mas yr$^{-1}$)} &
\colhead{(km s$^{-1}$)} &
\colhead{Type} &
\colhead{} }
\startdata
 15.5302& $-$0.5499& 18.18& $-$0.20& 11.64& 163&  344& $-$126& 282& DA& 1 \\
 45.8551& $-$8.1430& 18.74& $-$0.33& 10.50& 256&  191&   49& 239& DA& 2 \\
125.9965& 31.1984& 18.12& $-$0.64&  8.65& 310&  $-$16& $-$164& 242& DA& \\
134.6315& 42.4896& 18.53&  0.05& 12.74& 124& $-$229& $-$203& 180& DA& 3 \\
139.3937&  2.1568& 18.87& $-$0.54&  8.72& 430&  $-$98&  $-$41& 216& DA& 2 \\
140.1324&  2.9664& 18.55& $-$0.34& 11.11& 230&  $-$74& $-$219& 252&   & \\
151.1846&  8.6299& 18.06& $-$0.66&  8.57& 308& $-$173&  $-$59& 267& DA& \\
152.3864& 52.7773& 18.59& $-$0.48& 10.21& 286&  $-$56& $-$114& 172& DA& \\
156.3491&  0.7183& 18.44&  0.73& 14.54&  47&  $-$63&$-$1114& 248& DA& 4 \\
158.8389& 61.1174& 18.45& $-$0.62&  8.56& 377&  $-$40& $-$113& 214& DA& \\
159.9244& 46.2068& 18.81& $-$0.05& 12.43& 159& $-$224& $-$132& 196& DZ& 5 \\
161.4964& 59.0801& 17.76&  0.05& 12.70&  88&$-$1022&$-$1464& 746& DQ& 6 \\
166.8808& 48.9231& 19.47&  1.25& 15.09&  49& $-$726&  $-$79& 171& DC& 7,8 \\
170.0118& 44.4748& 18.56& $-$0.44& 10.48& 265&  $-$90&  $-$99& 168&   & \\
172.8554&  9.2366& 18.83& $-$0.64&  8.30& 449&  $-$69&   33& 163&   & \\
181.8450&  9.2898& 19.41& $-$0.24& 11.60& 294&  114& $-$260& 395&   & \\
184.3810& 61.0892& 18.06& $-$0.83&  6.20& 534&  $-$76&  $-$66& 255& DA& \\
190.1810& 67.1766& 18.25& $-$0.63&  8.44& 346& $-$183&  $-$92& 336&   & \\
193.7839& 46.9218& 19.19&  1.14& 14.97&  49&$-$1089& $-$114& 252&   & 8 \\
193.7839& 46.9218& 19.19&  1.14& 15.40&  38&$-$1089& $-$114& 197&   & 9 \\
194.4960& $-$2.2204& 19.15& $-$0.72&  7.69& 621&  $-$71&  $-$24& 220& DB& 2 \\
195.5034&  1.7721& 18.31& $-$0.47& 10.37& 240& $-$140&  $-$50& 169&   & \\
206.4670& 40.0171& 18.59& $-$0.63&  8.66& 396& $-$194&  $-$54& 378& DA& \\
216.7079&  9.8268& 17.89& $-$0.55&  9.69& 228&   47& $-$321& 351&   & 10 \\
222.2287&  5.3177& 18.00& $-$0.58&  8.64& 287& $-$132&  $-$16& 181&   & \\
223.1366& 10.8904& 18.74& $-$0.72&  7.74& 504&  $-$58&  $-$35& 162&   & \\
225.6319&  1.1794& 18.46&  0.39& 11.82& 174&  $-$79& $-$195& 174& DA+M& 2,11 \\
240.3110& 53.7709& 18.49&  0.14& 13.08& 104&   74& $-$385& 194& DA& 2 \\
243.6112& 38.5146& 19.22& $-$0.96&  2.38&1484&   14&  $-$41& 305&   & \\
245.7565& 30.5115& 17.91& $-$0.61&  9.14& 261&    3& $-$162& 201&   & \\
332.9242& 11.6012& 19.27& $-$0.13& 12.00& 229&   58& $-$165& 190&   & \\
342.5009& 12.6722& 19.50& $-$0.25& 11.27& 316&  130&  $-$16& 196& DQ& \\
355.2922&  0.5499& 19.14& $-$0.46& 10.39& 343&    9& $-$106& 173& DA& 2 \\
\enddata
\tablenotetext{1}{LP586$-$51; in Oppenheimer et al. (2001)}
\tablenotetext{2}{In Kleinman et al. (2004)}
\tablenotetext{3}{LP210$-$12}
\tablenotetext{4}{LHS 282}
\tablenotetext{5}{LP168$-$8}
\tablenotetext{6}{LHS 291}
\tablenotetext{7}{In L\'epine et al. (2003)}
\tablenotetext{8}{D assuming helium atmosphere}
\tablenotetext{9}{D assuming hydrogen atmosphere}
\tablenotetext{10}{LP500$-$28}
\tablenotetext{11}{LP621$-$58}
\end{deluxetable}

\clearpage


\begin{figure}
\plotone{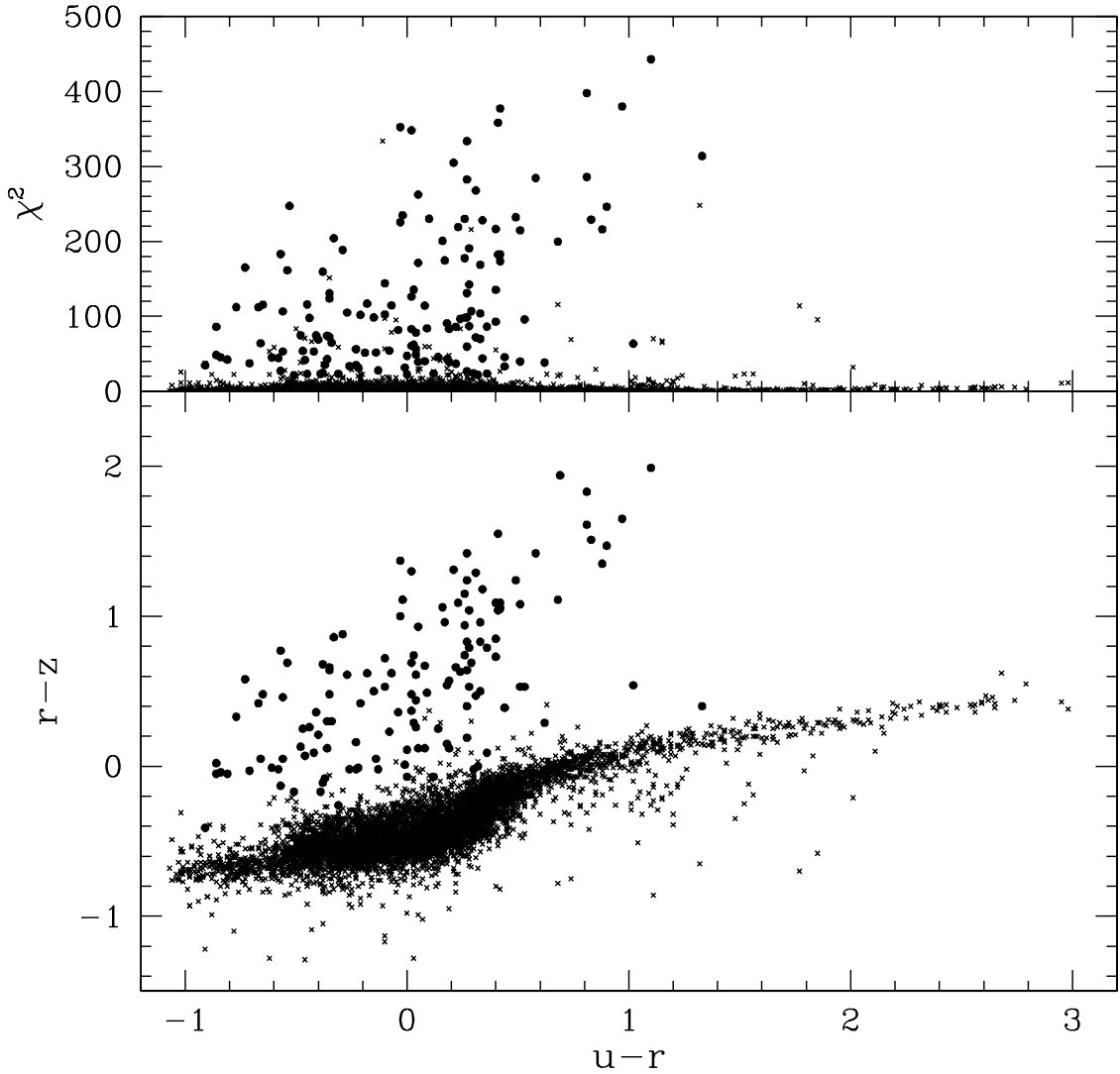}
\caption{Plots showing the identification of composite
WDs with cool companions, shown by filled circles.  The top panel
shows chi-square values when all five SDSS magnitudes are fit to
model colors for WD candidates with $V_{\rm tan} > 30$ km~s$^{-1}$.
The bottom panel shows a two-color diagram.
\label{Fig.1}}
\end{figure}

\clearpage
\begin{figure}
\includegraphics[scale=0.65,angle=-90] {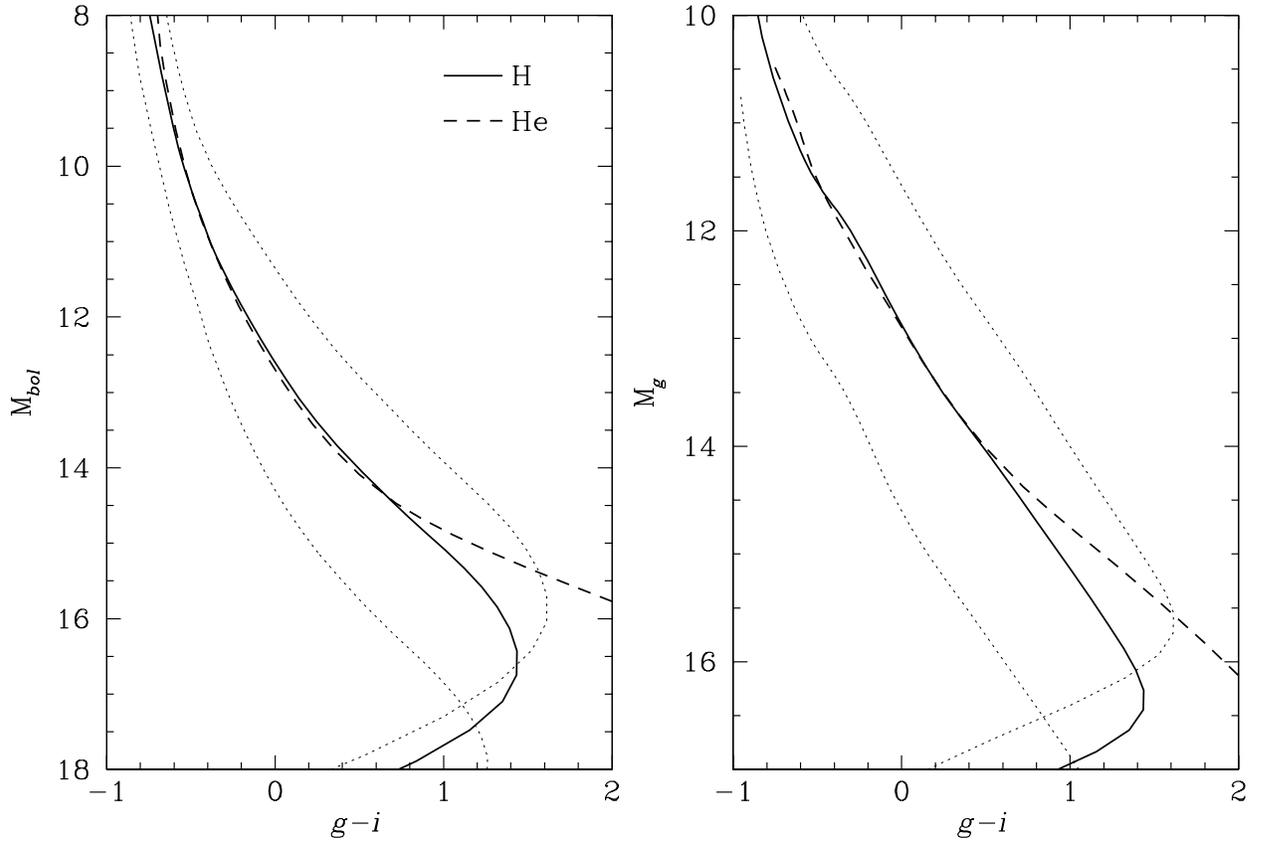}
\caption{Color-absolute magnitude relations for model WDs
with pure hydrogen (solid line) and pure helium (dashed line)
atmospheres with log~$g = 8$ from Bergeron et al. (1995) models.
Dotted lines show models with log~$g = 7$ for low mass WDs
(upper line) and log~$g = 9$ for high mass WDs (lower line).
\label{Fig.2}}
\end{figure}

\clearpage
\begin{figure}
\includegraphics[scale=0.65,angle=-90] {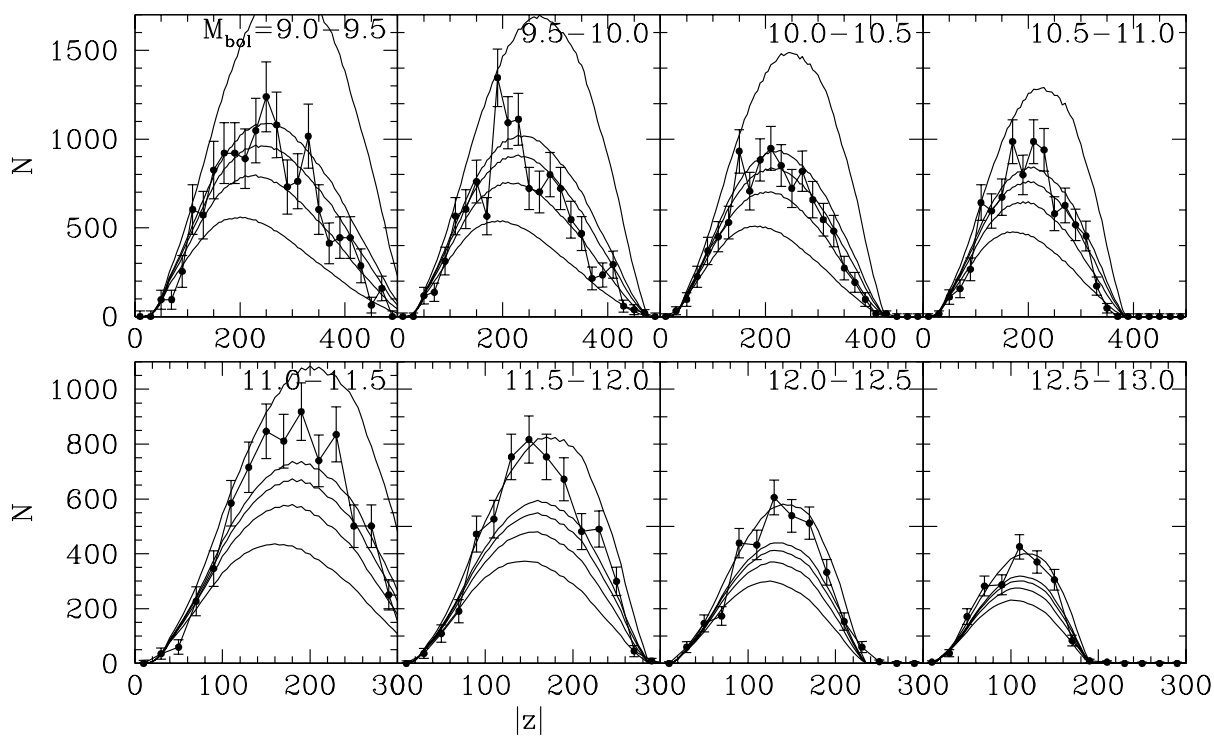}
\caption{The distribution of WD candidates with distance
from the Galactic plane for different luminosity bins.
The curves in each panel show models of the distribution
assuming disk scale heights of (from top to bottom)
$\infty$, 500, 400, 300, and 200 pc.
\label{Fig.3}}
\end{figure}

\clearpage
\begin{figure}
\plotone{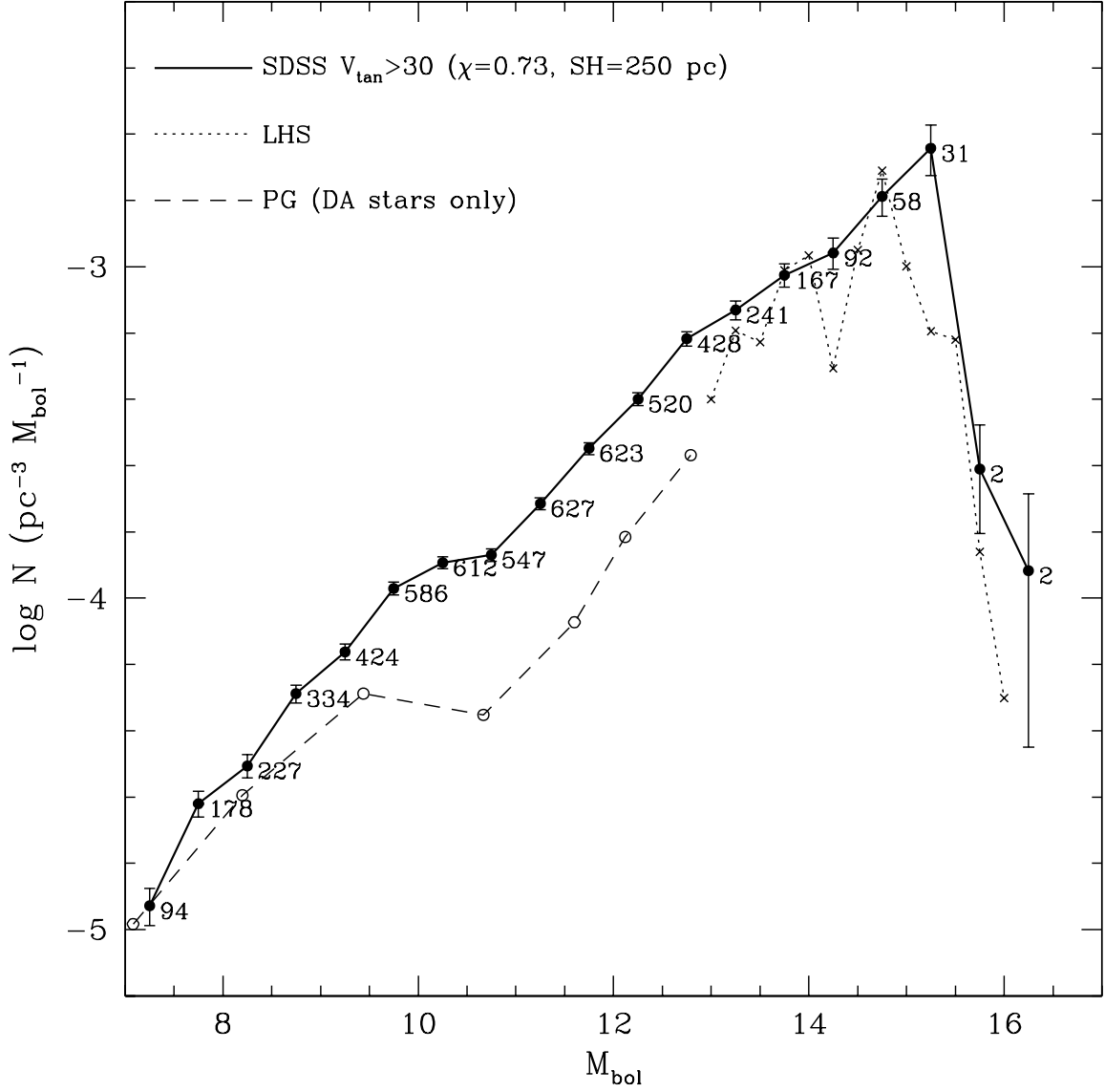}
\caption{The luminosity function derived in this paper.
The number of stars used for each data point is indicated.
Two results from the literature are shown for comparison:
the dotted line at the faint end is taken from Leggett et al.
(1998), based on the LHS Catalog; the dashed line at the
bright end is taken from Liebert et al. (2005), based on
analysis of the PG Survey, including DA WDs only.
\label{Fig.4}}
\end{figure}

\clearpage
\begin{figure}
\plotone{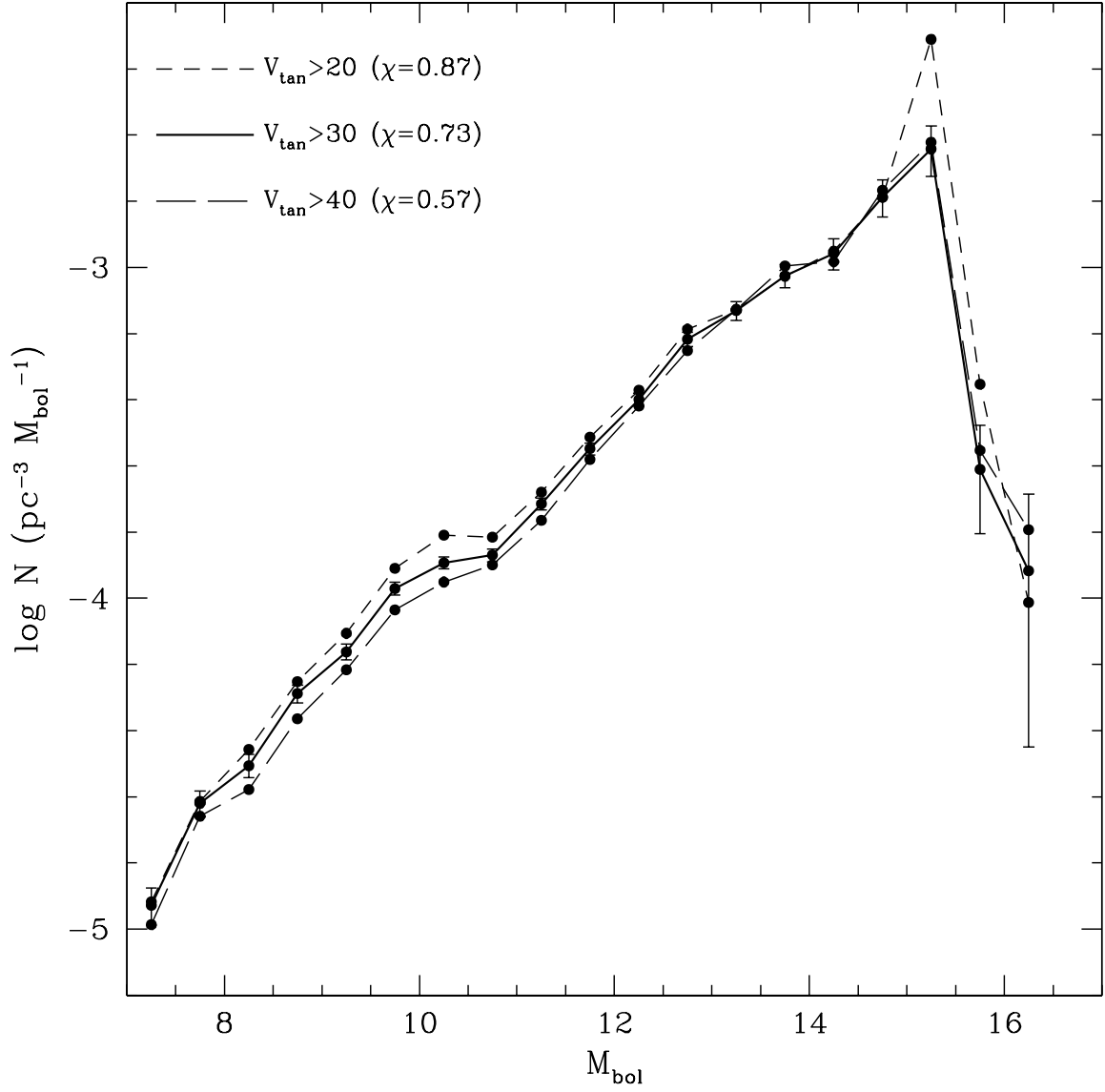}
\caption{The luminosity function derived using different
lower limits of tangential velocity.  The factor ($\chi$)
used to account for low velocity WDs below the selection limit
is described in the text.
\label{Fig.5}}
\end{figure}

\clearpage
\begin{figure}
\plotone{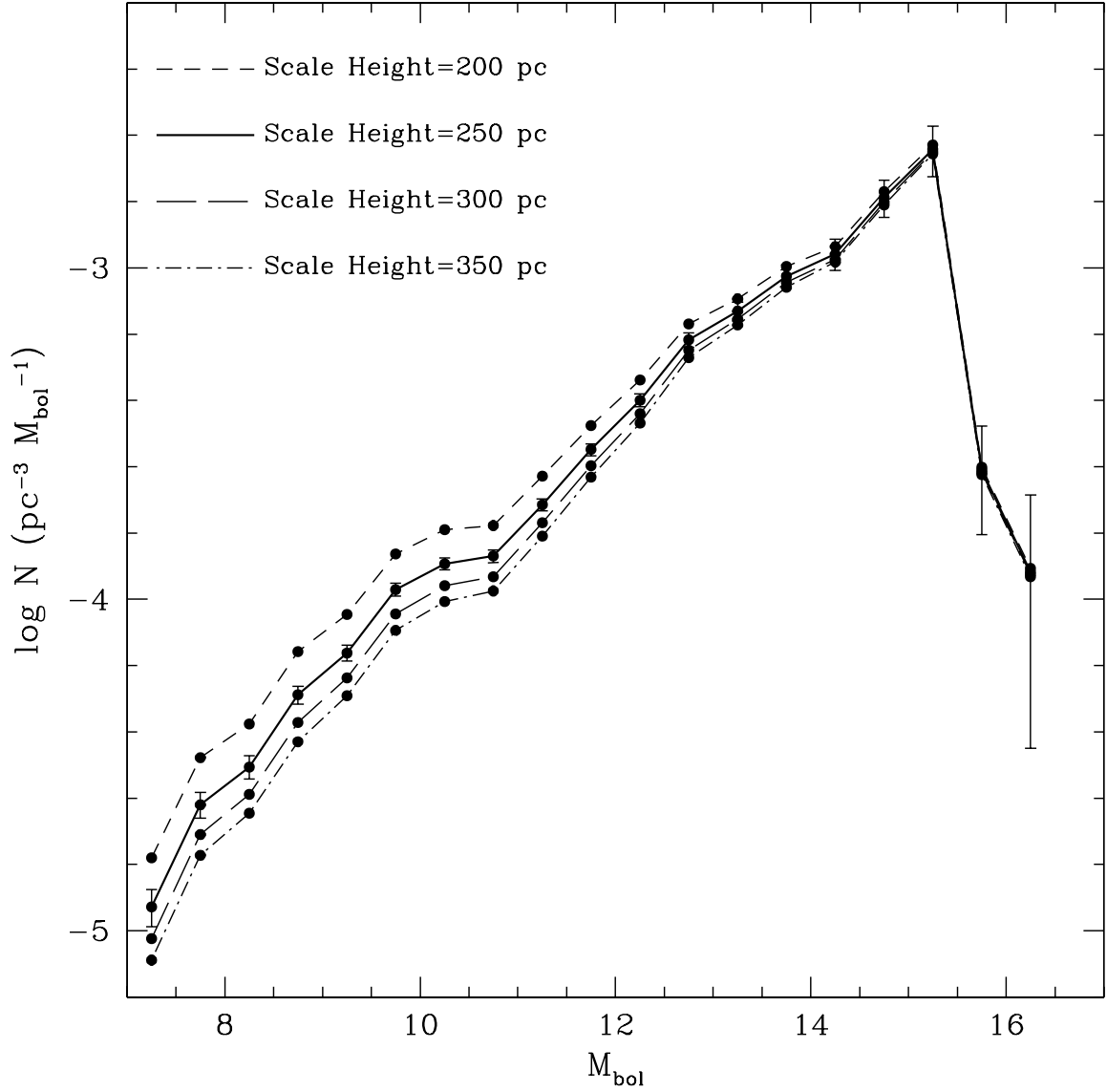}
\caption{The luminosity function derived using different
values for the scale height of the Galactic disk.
\label{Fig.6}}
\end{figure}

\clearpage
\begin{figure}
\plotone{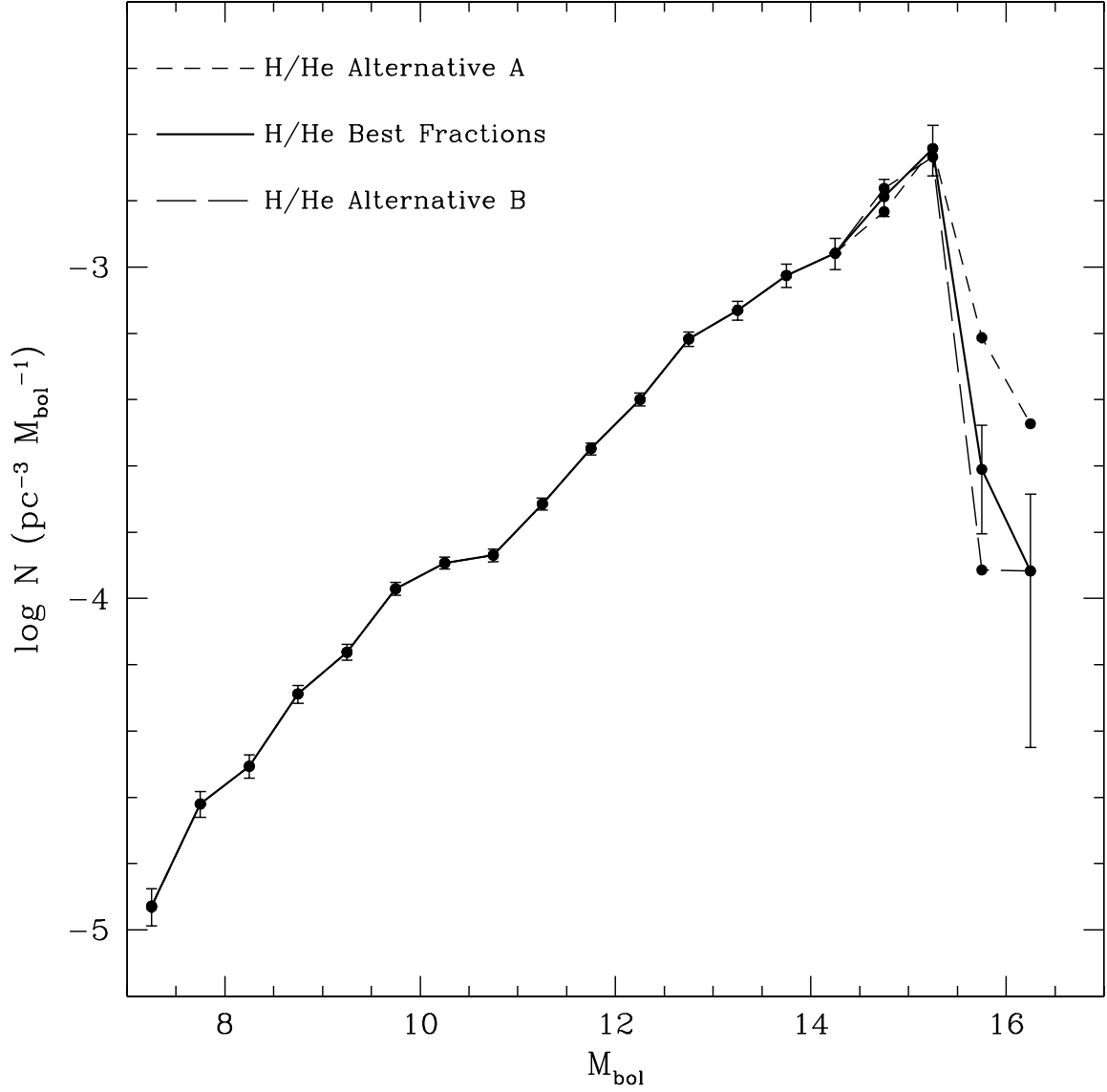}
\caption{The luminosity function derived assuming different
fractions of hydrogen- and helium-dominated atmospheres
for the coolest WDs.
\label{Fig.7}}
\end{figure}

\clearpage
\begin{figure}
\plotone{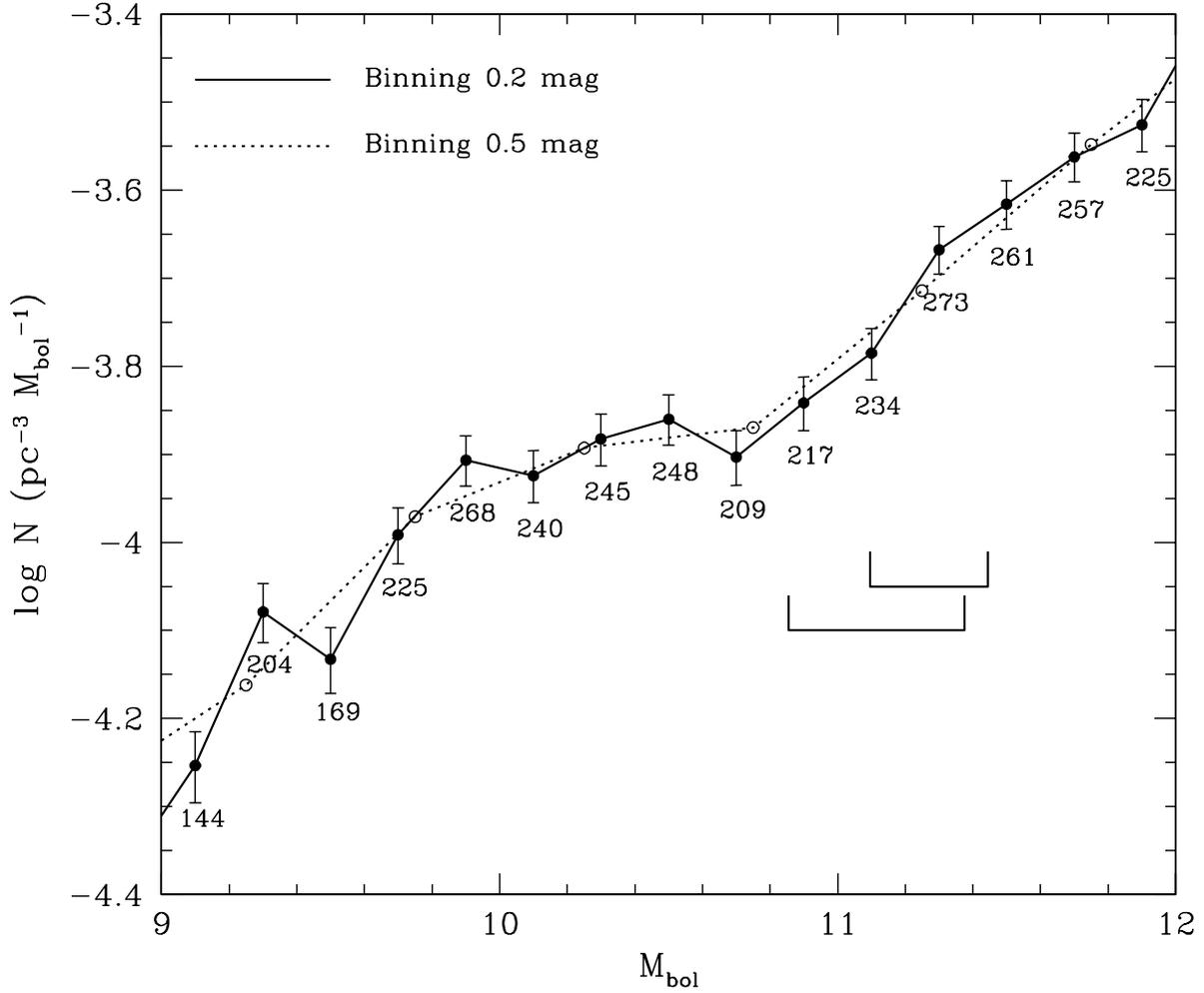}
\caption{A higher resolution plot of the middle portion of
Figure 4.  The two bars below the curve near $M_{\rm bol} = 11$
show the location of the ZZ Ceti instability strip observed by
Mukadam et al. (2004, top) and Bergeron et al. (2004, bottom).
\label{Fig.8}}
\end{figure}

\clearpage
\begin{figure}
\plotone{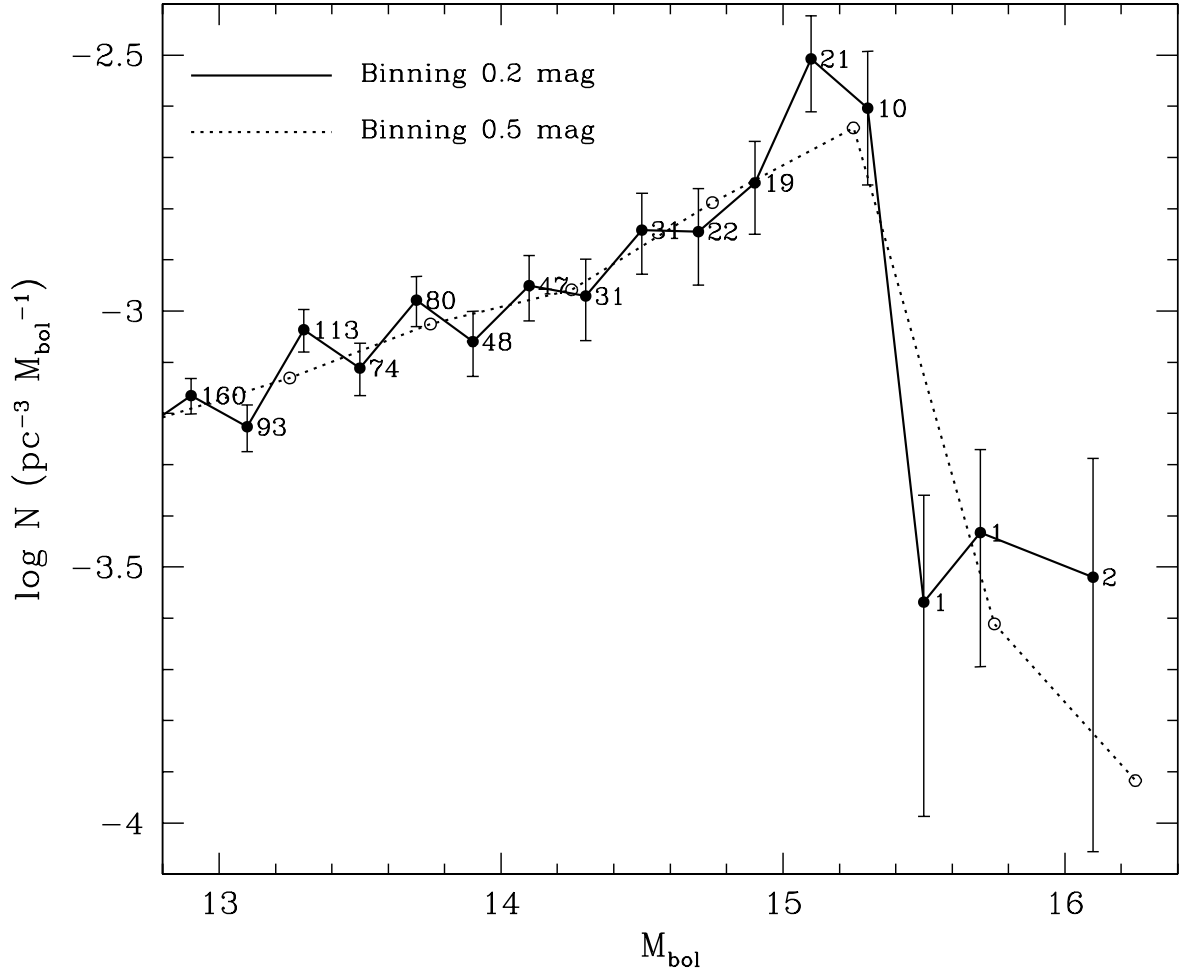}
\caption{A higher resolution plot of the low-luminosity end of
Figure 4.
\label{Fig.9}}
\end{figure}

\clearpage
\begin{figure}
\plotone{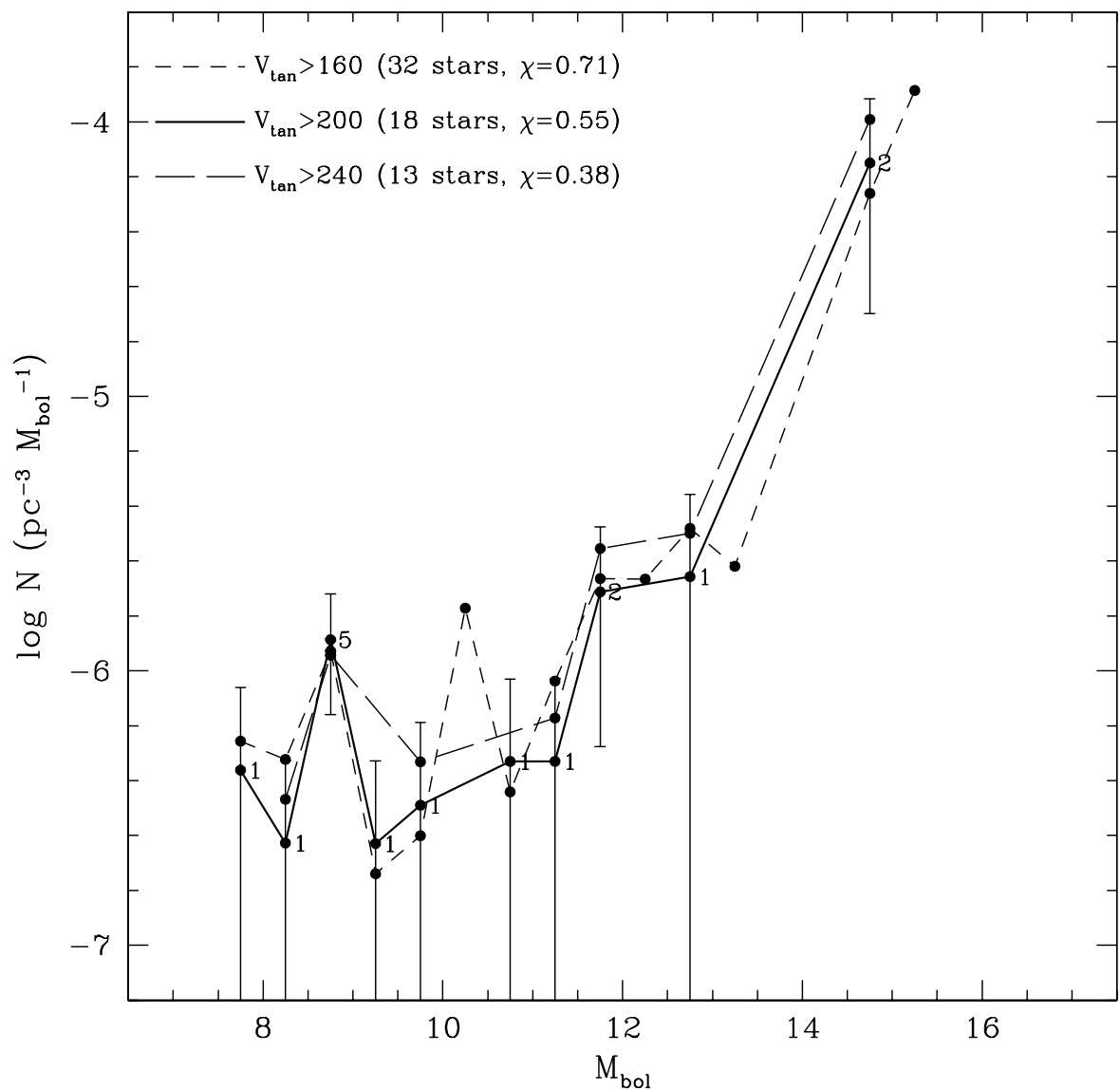}
\caption{A preliminary luminosity function for high-velocity WDs.
The three curves show the LF using different selection limits
of tangential velocity, and after correction for incompleteness
using the listed values of $\chi$.
The sample may be dominated by halo WDs, but probably includes
some fractions of disk, thick-disk, and high-mass WDs (see text).
\label{Fig.10}}
\end{figure}

\end{document}